\documentstyle[aps,preprint,pra]{revtex}
\begin{document}
\newcommand{\etal}{{\em et al.}\/}
\newcommand{\IP}{inner polarization}
\newcommand{\IPF}{\IP\ function}
\newcommand{\IPFs}{\IP\ functions}
\newcommand{\auth}[2]{#1 #2, }
\newcommand{\oneauth}[2]{#1 #2, }
\newcommand{\twoauth}[4]{#1 #2 and #3 #4, }
\newcommand{\andauth}[2]{and #1 #2, }
\newcommand{\jcite}[4]{#1 {\bf #2}, #3 (#4)}
\newcommand{\BOOK}[4]{{\it #1} (#2, #3, #4)}
\newcommand{\inbook}[5]{In {\it #1}, #2 (#3, #4, #5)}
\newcommand{\edit}[2]{Ed. #1, #2}
\newcommand{\erratum}[3]{\jcite{erratum}{#1}{#2}{#3}}
\newcommand{\JCP}[3]{\jcite{J. Chem. Phys.}{#1}{#2}{#3}}
\newcommand{\jms}[3]{\jcite{J. Mol. Spectrosc.}{#1}{#2}{#3}}
\newcommand{\jmsp}[3]{\jcite{J. Mol. Spectrosc.}{#1}{#2}{#3}}
\newcommand{\theochem}[3]{\jcite{J. Mol. Struct. ({\sc theochem})}{#1}{#2}{#3}}
\newcommand{\jmstr}[3]{\jcite{J. Mol. Struct.}{#1}{#2}{#3}}
\newcommand{\cpl}[3]{\jcite{Chem. Phys. Lett.}{#1}{#2}{#3}}
\newcommand{\cp}[3]{\jcite{Chem. Phys.}{#1}{#2}{#3}}
\newcommand{\pr}[3]{\jcite{Phys. Rev.}{#1}{#2}{#3}}
\newcommand{\PR}[3]{\jcite{Phys. Rev.}{#1}{#2}{#3}}
\newcommand{\PRL}[3]{\jcite{Phys. Rev. Lett.}{#1}{#2}{#3}}
\newcommand{\PRA}[3]{\jcite{Phys. Rev. A}{#1}{#2}{#3}}
\newcommand{\PRB}[3]{\jcite{Phys. Rev. B}{#1}{#2}{#3}}
\newcommand{\jpc}[3]{\jcite{J. Phys. Chem.}{#1}{#2}{#3}}
\newcommand{\jpca}[3]{\jcite{J. Phys. Chem. A}{#1}{#2}{#3}}
\newcommand{\jpcA}[3]{\jcite{J. Phys. Chem. A}{#1}{#2}{#3}}
\newcommand{\jcc}[3]{\jcite{J. Comput. Chem.}{#1}{#2}{#3}}
\newcommand{\molphys}[3]{\jcite{Mol. Phys.}{#1}{#2}{#3}}
\newcommand{\physrev}[3]{\jcite{Phys. Rev.}{#1}{#2}{#3}}
\newcommand{\mph}[3]{\jcite{Mol. Phys.}{#1}{#2}{#3}}
\newcommand{\cpc}[3]{\jcite{Comput. Phys. Commun.}{#1}{#2}{#3}}
\newcommand{\jcsfii}[3]{\jcite{J. Chem. Soc. Faraday Trans. II}{#1}{#2}{#3}}
\newcommand{\jacs}[3]{\jcite{J. Am. Chem. Soc.}{#1}{#2}{#3}}
\newcommand{\ijqcs}[3]{\jcite{Int. J. Quantum Chem. Symp.}{#1}{#2}{#3}}
\newcommand{\ijqc}[3]{\jcite{Int. J. Quantum Chem.}{#1}{#2}{#3}}
\newcommand{\spa}[3]{\jcite{Spectrochim. Acta A}{#1}{#2}{#3}}
\newcommand{\tca}[3]{\jcite{Theor. Chem. Acc.}{#1}{#2}{#3}}
\newcommand{\tcaold}[3]{\jcite{Theor. Chim. Acta}{#1}{#2}{#3}}
\newcommand{\jpcrd}[3]{\jcite{J. Phys. Chem. Ref. Data}{#1}{#2}{#3}}
\newcommand{\APJ}[3]{\jcite{Astrophys. J.}{#1}{#2}{#3}}
\newcommand{\astast}[3]{\jcite{Astron. Astrophys.}{#1}{#2}{#3}}
\newcommand{\arpc}[3]{\jcite{Ann. Rev. Phys. Chem.}{#1}{#2}{#3}}

\draft
\title{Electron affinities of the first- and second- row atoms:
benchmark ab initio and density functional calculations}
\author{Gl\^enisson de Oliveira and Jan M.L. Martin$^*$}
\address{Department of Organic Chemistry,
Kimmelman Building, Room 262,
Weizmann Institute of Science,
76100 Re\d{h}ovot, Israel. {\em Email:} \verb|comartin@wicc.weizmann.ac.il|
}
\author{Frank de Proft and Paul Geerlings$^*$}
\address{Eenheid Algemene Chemie (ALGC), Vrije Universiteit Brussel,
Pleinlaan 2, B-1050 Brussel, Belgium. {\em Email:}
\verb|fdeprof@vub.ac.be|}
\date{{\it Phys. Rev. A} MS\#AB7143; Received February 10, 1999; Revised April 22, 1999}
\maketitle
\begin{abstract}
A benchmark ab initio and density functional (DFT) study has been
carried out on the electron affinities of the first- and second-
row atoms. The ab initio study involves basis sets of $spdfgh$ and
$spdfghi$ quality, extrapolations to the 1-particle basis set
limit, and a combination of the CCSD(T), CCSDT, and full CI
electron correlation methods. Scalar relativistic and spin-orbit
coupling effects were taken into account. On average, the best ab 
initio results agree to better than 0.001 eV with the most
recent experimental results. Correcting for imperfections in the
CCSD(T) method improves the mean absolute error by an order of
magnitude, while for accurate results on the second-row atoms
inclusion of relativistic corrections is essential. The latter
are significantly overestimated at the SCF level; for accurate
spin-orbit splitting constants of second-row atoms inclusion
of (2s,2p) correlation is essential. In the DFT calculations
it is found that results for the 1st-row atoms are very sensitive
to the exchange functional, while those for second-row atoms
are rather more sensitive to the correlation functional.
While the LYP correlation functional works best for
first-row atoms, its PW91 counterpart appears to be preferable
for second-row atoms. Among ``pure DFT'' (nonhybrid)
functionals, G96PW91 (Gill 1996 exchange combined with Perdew-Wang
1991 correlation) puts in the best overall performance, actually slightly
better than the popular hybrid B3LYP functional. B3PW91 outperforms
B3LYP, while the recently proposed 1-parameter hybrid functionals
such as B1LYP seem clearly superior to B3LYP and B3PW91 for
first-row atoms. The best results overall are obtained with the
1-parameter hybrid modified Perdew-Wang (mPW1) exchange functionals
of Adamo and Barone [J. Chem. Phys. {\bf 108}, 664 (1998)], 
with mPW1LYp yielding the best results for first-row, 
and mPW1PW91 for second-row atoms.
Indications exist that a hybrid of the type
$a$ mPW1LYP+  $(1-a)$ mPW1PW91 yields better results than either
of the constituent functionals.
\end{abstract}

\section{Introduction}

The electron affinity (EA) of a system is the energy required for the 
reaction
\begin{equation}
A^- \rightarrow A + e^-
\end{equation}

Electron affinities have traditionally been regarded as one of the hardest
atomic or molecular properties to reproduce in an ab initio quantum
mechanical calculation. For starters, they
involve a change in the number
of valence electrons correlated in the system, and hence are very taxing tests for any
electron correlation method. In addition, they involve a pronounced change in the
spatial extent of the wave function, making them very demanding in terms of the
basis set as well.

The electron affinities
of the first-and second-row atoms have often been
used as benchmarks
\cite{Gda99,Gut98,Guo98,Wij98,Wij97,Eli97,Fis95,Ken92,Woo93,Nor91,Sun90,Nov91,Bau86a,Bau86b,Fel85,Rag85}
for high-level electronic
structure methods since (a) many of them are known experimentally to very high precision (e.g.\cite{CRC78});
(b) no such complications as geometry relaxation are involved; and (c) 
the computational demands required are still relatively modest.
Until recently, three of the first-and second-row atomic electron
affinities were imprecisely known experimentally (B, Al, and Si): this
situation was changed very recently by high-precision measurements in
recent experiments for B\cite{Sch98B}, Al\cite{Cal96,Sch98Al}, and Si\cite{Tho96}.

Density functional theory \cite{Par89,Dre90,Koh96} allows a cost-effective introduction of
electron correlation via the Kohn-Sham method\cite{Koh65} and the
use of exchange-correlation functionals. However, since the systematic extension of these
functionals towards the exact solution of the Schr\"odinger equations is not
possible hitherto, calculated results have to be compared with ab initio
wave function calculations or experiment in order to judge their
reliability and quality.
In recent years, many studies have evolved testing the
performance of density functional methods in the calculation of atomic
and molecular properties.
One of these properties which can be used to
critically test the available exchange-correlation functionals are
electron affinities.

DFT electron affinities have already
been obtained by a number of groups.
Pople et al. investigated the performance of the B-LYP exchange
correlation
functional in the calculation of atomization energies, ionization
energies, electron affinities and proton affinities using the 6-31G(d),
6-31+G(d), 6-311+G(2df,p) and 6-311+G(3df,2p) basis sets \cite{Gil92}.
In a test on the molecules of the well-known G2 thermochemical data
set\cite{Cur91}, a mean absolute deviation from experiment of
0.137 eV for the electron affinities (25 molecules) was found
for the largest basis set.
In a previous contribution\cite{Pro97}, two of us have studied ionization potentials and
electron affinities using the hybrid functionals B3LYP and B3PW91 and
Dunning's correlation consistent basis sets\cite{Dun89}.  For the largest basis set
studied (i.e. the AVTZ basis), a mean absolute deviation from
experiment of 0.13 eV for both of these functionals was found in the
calculation of electron affinities for the G2 set of molecules.
Schaefer and coworkers have studied electron affinities for a variety of
systems: sulphur fluorides\cite{Kin96a}, phosphorus fluorides\cite{Tsc96}, 
monochlorine fluorides\cite{Van96} and silicon fluorides \cite{Kin96b}. 
Galbraith and
Schaefer \cite{Gal96} also evaluated the electron affinities for F and
F$_{2}$ using a number of exchange-correlation functionals and the AVDZ,
AVTZ, AVQZ and AV5Z basis sets.  Moreover, they
studied the atomic electron affinities for the first row elements, 12
first row diatomic and 15 first row triatomic molecules using 6
different functional, amongst which some hybrid functionals\cite{Tsc97}.  
It was
found that for their series of tested molecules, the BLYP functional
provided the best agreement with experiment, the overall absolute error
being 0.21 eV.  For the B3LYP, BP86 and BHLYP functionals, the absolute
error lies around 0.3 eV, whereas the B3P86 and LDA errors are around
0.7 eV.
Recently, Curtiss et al\cite{Cur98} studied the performance of density 
functional methods in the calculation
of ionization energies and electron affinities on the so-called G2 ion
test set, which consists of the 63 atoms and molecules whose ionization
energies and electron affinities were included in the original G2 test
set, supplemented with 83 atoms and molecules.  Thus, they determined
the performance of the seven exchange correlation functionals in the
calculation of 58 electron affities.  It was conclude that for this set
and the 6-311+(3df,2p) basis set, the mean absolute deviations were
0.697 eV (LDA), 0,113 eV (BLYP), 0.121 eV (BPW91), 0.193 eV (BP86),
0.131 eV (B3LYP), 0.145 eV (B3PW91) and 0.596 eV (B3P86).

The purpose of the present work is twofold. First of all, we will
try to establish whether present-day state-of-the-art wavefunction
based methods will consistently yield `the right result for the
right reason'. As a by-product, we will obtain basis set limit values
for the nonrelativistic, clamped-nuclei electron affinities, which
will serve for the second purpose.
This involves the testing of the performance and basis set dependence of
different exchange-correlation density functionals in the calculation of
these electron affinities.

\section{Computational methods}

\subsection{Density functional calculations}

Density functional calculations were performed using Gaussian 94\cite{g94}
running on the Cray J916/8-1024 of the Brussels Free Universities Computer Centre,
and Gaussian 98\cite{g98} running on the SGI Origin 2000 of the Faculty of
Chemistry at the Weizmann Institute of Science.

In order to account for possible errors in the numerical integration due
to the diffuseness of the charge density, in particular of course for the
anions, and the high angular momentum in the basis set, a fine grid of
590 angular Lebedev nodes and 99 radial nodes was used
and tightened convergence criteria for
the Kohn-Sham equations were specified, such that the
tabulated results for the electron
affinities can be considered precise to 10$^{-4}$ eV.
\typeout{Let op het verschil tussen ``precisie'' en ``accuraatheid''!}

A wide variety of exchange-correlation functionals $E_{xc}$ was considered.
Among the ``pure DFT'' functionals, these are the following:
\begin{itemize}
\item{The Local Density Approximation (LDA), which actually uses
Slater's expression for exchange (S) \cite{Sla51} and Vosko, Wilk and
Nusair's expression
for the correlation energy of the uniform electron gas \cite{VWN},
parameterized
using Ceperley and Alder's quantum Monte-Carlo results\cite{Cep80};}
\item{The gradient corrected B-LYP,B-P86 and B-PW91
functionals, which are combinations of Becke's
1988 (B88, or simply B)
gradient-corrected exchange functional\cite{Bec88}
with correlation functionals due to Lee, Yang, and Parr
(LYP)\cite{Lee88}, Perdew (P86)\cite{Per86}, and
Perdew and Wang (PW91)\cite{Per92}, respectively;}
\item{The combination of the PW91 correlation functional with
the exchange functional proposed in the same paper\cite{Per92},
a combination usually denoted by the acronym GGA91 for Generalized
Gradient Approximation-1991}
\item{Combinations of the LYP and PW91 correlation functionals with
the 1996 exchange functional proposed by Gill\cite{Gil96}, denoted
G96LYP and G96PW91, respectively}
\end{itemize}

Among ``hybrid'' functionals (i.e. those having a nonzero coefficient
for the true Hartree-Fock exchange $E_{x}^{HF}$) we have considered the
following:
\begin{itemize}
\item The popular B3LYP\cite{Bec93,Ste94} functional, which takes the form
\begin{equation}
E_{xc} = a_{x0}E_{x}^{LDA} + (1-a_{x0}) E_{x}^{HF}
+a_{x1}\Delta E_{x}^{B88} + (1-a_{c}) E_{c}^{LDA} + a_{c}E_{c}^{LYP}
\end{equation}
in which the three constants $a_{x0}=0.80$, $a_{x1}=0.72$, $a_{c}=0.81$ were
originally empirically
determined by Becke\cite{Bec93} using the P86 correlation functional,
and the implementation in the Gaussian series of programs\cite{Ste94}
uses the VWN functional 3\cite{VWN} for $E_{c}^{LDA}$ rather than the
VWN functional 5 employed by Becke\cite{Bec93}
\item the B3PW91 functional, which has the same form as the
B3LYP functional except that $E_{c}^{PW91}$ is used instead of
$E_{c}^{LYP}$
\item a number of the new one-parameter hybrid functionals proposed
by Adamo and Barone
\begin{equation}
E_{xc} = a_{0}E_{x}^{HF} + (1-a_{0}) (E_{x}^{LDA}+\Delta E_{x}^{GC})
+ E_{c}
\end{equation}
in which $\Delta E_{x}^{GC}$ is some gradient correction to the
exchange functional (e.g. B88, PW91, G96), and $E_{c}$ represents
any suitable
correlation functional.  From an analysis\cite{PEB} based on perturbation
theory, $a_{0}$ takes the nonempirical
value $1/4$. In the present paper, we have considered B1LYP and
B1PW91\cite{B1LYP} (i.e. the 1-parameter analogs of B3LYP and
B3PW91), as well as the newer LG1LYP functional\cite{LG1LYP} which uses
the Lacks-Gordon\cite{LG} expression for $E_{x}^{GC}$ and the
mPW1LYP and mPW1PW91 functionals\cite{MPW}, in which the nonlocal
exchange is given by a modification of $E_{x}^{PW91}$ for better
treatment of long-range interactions (the small density, large
gradient regime).
\end{itemize}

\subsection{Ab initio calculations}

The CCSDT (coupled cluster with
all single, double, and triple excitations\cite{ccsdt})
calculations were carried
out using ACES II\cite{aces} running on a DEC Alpha 500/500
workstation at the Weizmann Institute of Science; all other
ab initio calculations reported in this work were carried out
using MOLPRO 98.1\cite{m98} running on a Silicon Graphics Octane
workstation at the Weizmann Institute.

The valence calculations were carried out using the
augmented correlation-consistent valence $n$-tuple zeta
(aug-cc-pV$n$Z, or AV$n$Z for short) basis sets of Dunning and
coworkers\cite{Ken92}. The contracted sizes for the various AV$n$Z
basis sets for [second-row/first-row/hydrogen] atoms are as follows:
AVDZ [5s4p2d/4s3p2d/3s2p], AVTZ [6s5p3d2f/5s4p3d2f/4s3p2d],
AVQZ [7s6p4d3f2g/6s5p4d3f2g/5s4p3d2f],
AV5Z [8s7p5d4f3g2h/7s6p5d4f3g2h/6s5p4d3f2g]; in addition,
for first-row and
hydrogen atoms only, we considered
AV6Z [8s7p6d5f4g3h2i/7s6p5d4f3g2h].

Except where indicated otherwise, ROHF (restricted open-shell
Hartree-Fock) reference wave functions were used throughout.

The SCF component of the total energy was extrapolated using a
geometric expression\cite{Fel92} of the type $A+B/C^{n}$ applied to
AV$n$Z energies with $n$=Q, 5, 6 for first-row atoms and
$n$=T, Q, 5 for second-row atoms. The CCSD(T) (coupled cluster
with all single and double excitations and a quasiperturbative
treatment of connected triple excitations\cite{Pur82,Rag89,Wat93})
valence correlation
energy was extrapolated using both the 3-parameter expression
$A+B/($n$+1/2)^{\alpha}$ proposed by one of us\cite{l4} and the
two-parameter expression $A+B/n^{3}$ proposed by Helgaker and
coworkers\cite{Hal98}: both expressions are based on the known
asymptotic convergence behavior\cite{Sch63,Kut92} of pair
correlation energies as a function of the maximum angular momentum
present in the basis set.

Imperfections in the treatment of connected triple excitations
are corrected by means of CCSDT calculations in the AVQZ basis set.
Finally, the effect of connected quadruple and higher excitations
is approximated by full configuration interaction (FCI) in the largest
basis set where this is feasible with the Knowles-Handy\cite{Han85}
determinantal code. For B and Al, this is AVQZ; for C and Si, AVTZ;
for the other elements AVDZ.

The effect of inner-shell correlation was determined as the
difference between valence-only and all-electron CCSD(T) calculations
using the Martin-Taylor\cite{hf,cc} family of core-correlation basis sets.
The MTavqz basis set corresponds to a completely uncontracted
AVQZ basis set augmented with $1p3d2f$ high-exponent functions of which
the exponents are obtained by successively multiplying the largest
exponent already present in that angular momentum with a factor of 3.0.
The MTav5z basis set was similarly obtained from the AV5Z basis set
but with $1p3d2f1g$ high-exponent functions added.

Scalar relativistic effects were approximated by the first-order
perturbation correction\cite{Cow76,Mar83} of the Darwin and
mass-velocity (DMV) terms. For technical reasons, these calculations
were carried at at the ACPF (averaged coupled pair
functional\cite{Gda88}) level. Since great flexibility in the $s$ and
$p$ functions is essential for this type of effect, we employed the
MTavqz basis set throughout for this contribution.

Spin-orbit coupling constants were evaluated at the CASSCF-CI level
using the $spdf$ part of the MTav5z basis set. (For a recent review 
of the methodology involved, see Ref.\cite{Pey95}.)

\section{Results and discussion}

A summary of our computed results and their different components
is presented in Table I together with the experimental
results, while a selection of previously computed literature
values is presented in Table II.

\subsection{First-row atoms}

An indication for the error introduced by our use of finite basis
sets and extrapolations can be obtained from our results for the
EA of hydrogen atom, for which the computed results represent exact
solutions within the respective finite basis sets.

The three-point geometric extrapolation for the SCF component adds
only about 0.0001 eV to the largest-basis set (AV6Z) result. The
2-point Halkier extrapolation, however, still adds some 0.0041 eV to
the final result. The latter, 0.75416
eV, agrees excellently with the most precise
measurement, 0.754195(19) eV\cite{CRC78}; the somewhat higher
observed value for deuterium, 0.754593(74) eV\cite{CRC78}, suggests
that deviations from the Born-Oppenheimer approximation (not
considered in the present work) may account for 0.0002--0.0004 eV;
hence this is probably a more realistic assessment of the residual
error in our calculation than the difference of 0.00004 eV between
computed and observed EAs.  The only term other than SCF and
valence correlation which contributes to our computed result is an
essentially negligible (4$\times10^{-5}$ eV) contribution of Darwin
and mass-velocity (DMV) effects.

The electron affinity of the boron atom, imprecisely known for a long
time, was very recently redetermined to high accuracy by Scheer
et al.\cite{Sch98B} as 0.279723(25) eV, in perfect agreement with a
very recent relativistic coupled cluster calculation by Eliav et al.
\cite{Eli97} in an exceedingly large $[35s26p20d14f9g6h4i]$ basis set,
as well as the numerical relativistic MCSCF calculation by
Fischer et al.\cite{Fis95}.
Our present best calculated result, 0.27858 eV, meets the
0.001 eV accuracy target using no larger basis sets than
$[8s7p6d5f4g3h2i]$. Again, the basis set extrapolation beyond AV6Z
amounts to essentially nil for the SCF contribution but 0.004 eV for
the valence correlation energy. The $n$-particle space calibration, in
this case, was carried out at the FCI/AVQZ level, and amounts to no
less than 0.0191 eV --- about three-quarters of which consists of
imperfections in the treatment of connected triples. As a more
extreme case of a general trend, the results reflect imbalance
between the quality of the CCSD(T) treatment for neutral and anion
--- in this case close to exact for B but rather less so for
B$^{-}$. Inner-shell correlation increases EA by 0.0043 eV,
while DMV effects reduce EA by 0.0013 eV and spin-orbit effects by
another 0.0006 eV.

Our best calculation for carbon, 1.26298 eV, agrees to within
experimental uncertainty with the experimental value 1.2629(3) eV.
The amounts bridged by the extrapolation parallel those found for H
and B. $n$-particle calibration accounts for 0.013 eV, split about
2:1 between imperfections in the treatment of connected triples and
effects of connected higher excitations.
Spin-orbit and scalar relativistic effects lower the EA by 0.003
eV each. Inner-shell correlation has the highest contribution of the
first-row atoms, 0.007 eV.

Nitrogen atom has no bound anion. For oxygen, our best calculation is
within 0.0005 eV of the very precisely known experimental value. In this
case, extrapolation even from the AV6Z basis set contributes a solid
0.016 eV to the final result --- it should be noted that the valence
correlation component of EA is almost three times larger in absolute
value than that in C. While the spin-orbit contributions largely
compensate between neutral and anion (reflected in the fairly small
EA contribution of -0.002 eV), the DMV contribution is 
relatively important at -0.006 eV (as expected). 
The $n$-particle correction, at
0.012 eV, largely consists of effects of connected quadruple and
higher excitations --- the difference between CCSDT and CCSD(T) only
amounts to about 0.002 eV.

The EA for F has traditionally been known as one of the very hardest
quantities to reproduce from a theoretical calculation. Our calculated
value is 0.004 eV higher than the experimental result --- which is
still close to 0.1\% accuracy relatively speaking. The basis set
extrapolation covers similar amounts as in O, while both the spin-orbit
(-0.016 eV) and DMV (-0.009 eV) contributions are quite sizable.
Inner-shell correlation contributes a similar amount as in B. The main
uncertain factor in our calculation is the deceptively small
$n$-particle calibration contribution of 0.0006 eV, which is actually
the result of a cancellation between imperfections in the CCSD(T)
treatment of connected triple excitations (-0.009 eV) on the one hand,
and the effect of connected higher excitations (+0.010 eV) on the
other hand. Unfortunately the largest basis set in which we could
carry out FCI calculations was AVDZ, but we expect the FCI-CCSDT
difference to converge as fast as in the case of B or C. The
considerable basis set variation of the CCSDT-CCSD(T) differences as
well as the clear downward trend,
progressing from +0.003 eV (AVDZ) over -0.006 eV (AVTZ) to -0.009 eV
(AVQZ) strongly suggests that this difference would be substantially
larger near the $n$-particle basis set limit. If we assume an
$A+B/l^{3}$ extrapolation for this difference (equivalent to carrying
out a valence correlation extrapolation on CCSDT rather than CCSD(T) values
for this case) we obtain a further lowering by 0.002 eV, bringing the
calculated EA down to 3.4031 eV, within 0.002 eV of the experimental
value of 3.401190(4) eV.

Of previously computed results for F, Gutsev et al. (CCSDT/AV5Z,
3.395	eV) and Curtiss et al. (G3 theory, 3.400 eV) are both in
excellent agreement with experiment. However, the former includes
neither spin-orbit nor DMV contributions, and their inclusion would
reduce the result to 3.370 eV. The G3 value does include spin-orbit terms
(experimentally derived) but not DMV, and would be reduced to 3.391
eV upon inclusion of the latter.

\subsection{Second-row atoms}

The previously rather imprecisely known\cite{CRC78} EA of aluminum was very
recently redetermined. Calabrese et al.\cite{Cal96} obtained 0.44094(66)
eV, while Scheer et al.\cite{Sch98Al} obtained the more precise, and
substantially lower value 0.43283(5) eV. Our own calculations agree to
four figures with this latter value. Extrapolation of the valence
correlation contribution beyond AV5Z accounts for only 0.0036 eV,
while inclusion of inner-shell correlation lowers EA by 0.016 eV,
almost perfectly cancelling the increase of 0.015 eV from
$n$-particle correction. As in isovalent B, imperfections in the
treatment of connected triples in CCSD(T) make up the bulk of that
effect. Spin-orbit coupling and scalar relativistic effects weigh
in at -0.0038 and -0.0054 eV, respectively.

The EA of Si was very recently revised to 1.38946(6) eV by Thogersen
et al. \cite{Tho96}. Our own calculation comes within 0.001 eV of
that value. With a substantial spin-orbit splitting in Si($^{3}P$) and
none at all in Si$^{-}$($^{4}S$), we find the spin-orbit
contribution to EA to be the second-largest
of the atoms surveyed, -0.018 eV, while scalar relativistic
effects are less substantial at -0.008 eV. Basis set extrapolation
bridges 0.006 eV in this case; inner-shell correlation is less
prominent than in Al but still affects the result by -0.010 eV,
which interestingly again nearly cancels the $n$-particle calibration
correction. The latter is about evenly split between imperfections in
the treatment of connected triple excitations and the effects of
connected quadruple and higher excitations.

In the final three atoms, basis set convergence appears to be
particularly slow, as witnessed by the fact that extrapolations from
AVQZ and AV5Z results cover 0.023, 0.026, and 0.030 eV, respectively,
for P, S, and Cl. When using AV5Z and AV6Z results for Cl, some 0.017
eV is still bridged. Under these circumstances, it is not surprising
that accuracy would be somewhat lower; and indeed, our computed results
for P, S, and Cl are too low by about 0.002 eV on average.

Given how diffuse particularly the P anion is (the isovalent N anion
is not even bound), one might wonder whether even the AV$n$Z basis
sets are sufficiently saturated in the anion region. In an attempt
to establish this, we have carried out calculations for P, S, and Cl
using dAV$n$Z (doubly-augmented V$n$Z) basis sets, in which the
additional set of diffuse functions was generated simply by
multiplying the lowest exponents already present by 0.25.
Particularly for P, but 
less so for S and Cl, there is a nontrivial difference
between AVQZ/AV5Z and dAVQZ/dAV5Z extrapolated limits: 0.0021 eV for P,
and 0.0011 eV for S and Cl. This leads to revised values that are
in perfect agreement with experiment for P and S, while the revised result
for Cl is only 0.0015 eV too low.

Aside from these specific remarks, we can make some general observations.

First of all, the mean absolute deviation between our best computed
ab initio values and the most recent experimental values is only
0.0009 eV, with the largest individual error, 0.0018 eV, seen for P.
To the best of our knowledge (see Table II),
this level of accuracy is unprecedented in the literature for this
property.

The inclusion of corrections for imperfections in the CCSD(T)
method is absolutely indispensable for this level of accuracy:
neglecting them raises the mean absolute error by more than
an order of magnitude, to 0.009 eV. This contribution, as noted
above, is generally dominated by corrections for imperfections in
the treatment of connected triple excitations, i.e. the difference
between CCSD(T) and CCSDT.

The contribution of inner-shell correlation stabilizes the anion over
the neutral in the first-row atoms: in absolute value, it goes through
a maximum for C although in relative terms, it monotonically decreases
in importance from left to right in the periodic table. For second-row
atoms, core correlation stabilizes the neutral over the anion, and
monotonically decreases from left to right in the Periodic Table.

As expected, the contribution of scalar relativistic
(Darwin and mass-velocity, DMV) effects mounts
from left to right within each row, and is more important for the
second row than for the first row. As seen in Table III,
our relativistic contributions
follow the same trends as those obtained in the numerical SCF
calculations of Garc{\'\i}a de la Vega\cite{Gar95} and of
Koga et al.\cite{Tat97}, particularly the consistent favoring
of the more compact neutral atom over the more diffuse anion.
However, in absolute value our ACPF/MTav5z calculated DMV
contributions are systematically
smaller than the numerical HF results; the difference increases from
left to right in the Periodic Table and becomes fairly substantial for
F and Cl. As is readily seen by comparing SCF/MTav5z and ACPF/MTav5z
results, this mostly reflects the effect of electron correlation on
the correction, which one would intuitively expect to decrease the
effect of a one-electron property that is most important for the
inner-shell electrons. Comparison of MTavqz and MTav5z results reveals
that our computed contributions are converged in terms of the basis
set to $\leq5\times10^{-5}$ eV
at the ACPF level and $\leq10^{-5}$ eV at the SCF level. The small difference
between the present SCF level contributions and the numerical HF results
reflects the inclusion of some additional scalar relativistic effects in the
latter, particularly the two-electron Darwin term which we did not consider.
Evidently their importance, at the Hartree-Fock level,
mounts from $-$0.00004 eV for B or $-$0.0001 eV for Al to about $-$0.0008--9
eV for F and Cl. It is not a priori clear how electron correlation would
affect these contributions, although a reduction in importance would
seem plausible.

The spin-orbit contributions likewise mount from left to right and
from top to bottom in the Periodic Table; however, because such
systems as C$^{-}$, Si$^{-}$, and P do not exhibit any first-order
spin-orbit splitting, the contributions to EA at first sight seem more
erratic.

To the accuracy relevant here, it hardly appears to matter whether the
observed or the best computed fine structures are used for calculating
the spin-orbit contribution. As seen in Table IV, the computed values
are clearly near convergence with respect to the basis set.
For the first-row atoms, the CASSCF values are quite close to
experiment but this holds much less true for the second-row atoms.
Inclusion of external valence correlation usually seems to lower the
computed values and bring them away from experiment, while the
inclusion of $(2s,2p)$ correlation for the second-row atoms leads to
a dramatic improvement in the quality of the results. Inclusion
of correlation from the deep-lying (1s) orbitals has little effect on
the second-row results, as expected, but for first-row atoms a
somewhat greater contribution is seen.

Of the previous calculations summarized in Table II, the one systematic
study that most
closely reproduces our present benchmark values are the very recent
benchmark calculations of Gdanitz\cite{Gda99}, which were carried out
using a variant of the multireference ACPF\cite{Gda88} method involving explicit
interelectronic distances, MRACPF-$r_{12}$\cite{Gda93}. (In fact, since
the author of Ref.\cite{Gda99} was apparently unaware of the revised
experimental EA of B\cite{Sch98B}, his accuracy for B is better than claimed in Ref.\cite{Gda99}.)
Nevertheless,
even using this elaborate method, the errors in the O and F electron
affinities obtained in that work\cite{Gda99} are still an order of
magnitude larger than those in the present work. Part of the
discrepancy is due to the reliance, for relativistic corrections,
on the numerical Hartree-Fock values of Garc{\'\i}a de la
Vega\cite{Gar95}, which we have seen above to be an overestimate
for the scalar relativistic contribution.

\subsection{Density functional results}

The suitability of DFT methods for calculating electron affinities has
been the subject of some debate in the literature. It was noted early
\cite{Sho77} that numerical LDA calculations on H$^-$ do not yield a
bound HOMO, and hence no electron affinities can be obtained unless
the anion is artificially stabilized by a Watson-sphere potential.\cite{Bec93}
The Schaefer group\cite{Gal96,Tsc97}  (see also Ref.\cite{Pro97})
however carried out EA calculations
using finite Gaussian basis sets with a variety of exchange-correlation
functionals and found quite reasonable agreement with experiment. These
at first sight contradictory findings were largely reconciled by
R\"osch and Trickey\cite{Ros97}, whose arguments will be briefly summarized
here.

The main cause of the problem is the fact that the spurious self-repulsion
of the electron in the Coulomb potential is not exactly canceled by the
corresponding term in the (approximate) exchange potential. (Exact
cancellation occurs both for Hartree-Fock exchange and for the
exact Kohn-Sham potential.\cite{Trick}) This `self-interaction error'
results in 
an exchange-correlation potential which for large $r$ approaches zero 
rather than the correct limit $-1/r$\cite{Per80}.
As a result, Kohn-Sham orbital energies are 
artificially shifted upward by amounts on the order of several eV; while
this is a mere annoyance for calculations on neutral systems, this is on the
same order of magnitude as the highest occupied orbital energies in anions and
leads to the latter becoming positive. Such an orbital (in an infinite
basis set) is non-normalizable, and in fact corresponds to a combination
of incoming and outgoing scattered waves. However, a finite basis set of
Slater or Gaussian basis functions, no matter how diffuse or extended,
will in effect confine the orbital to a finite-sized sphere and thus
render it artificially normalizable. The question as to whether a DFT
calculation of the electron affinity as $E(A)-E(A^-)$
will yield an acceptable result then largely hinges on whether the
artificially normalizable orbital itself will be significantly affected
by the incorrect asymptotic shape of the approximate potential, as well
as how well the self-interaction error in the total energy cancels between 
neutral and anionic species. 

The self-interaction error is mitigated by
the use of `hybrid' functionals like B3LYP or mPW1PW91, since the
Hartree-Fock component is free of self-interaction. (In the present work,
we found that all hybrid functionals yielded all-negative occupied orbital
energies for Cl$^-$, as did all hybrid functionals other than B3PW91
for F$^-$. The other anions still exhibit positive highest occupied orbital
energies.)  
In addition, a rigorous self-interaction correction (SIC)\cite{Per81}
can be introduced,
at the expense of introducing orbital-dependent potentials and orbital
representation invariance problems\cite{Ped85}. 
Numerical orbital calculations with
such self-interaction corrected DFT
methods (e.g.\cite{Gra95,Che96}) yield agreement with experiment for atomic
electron affinities comparable to that with the better hybrid functionals
studied here. (However, initial SIC-LDA results\cite{Goe97} for molecular binding energies 
and geometries were in fact poorer than with standard LDA, and even
more so compared with generalized gradient approximations.)

We note that the prime application for DFT calculations of electron
affinities would be large molecules where no other approach is
currently computationally feasible, and that the spatial extent of
the charge distribution in such systems would help reduce the self-repulsion
error\cite{Ros97}. By comparison, atoms represent a `worst-case scenario', so it would
definitely be of interest to know if accurate atomic EAs can be obtained
at all using modern `pure DFT' and `hybrid' exchange-correlation functionals
and finite basis sets. We will demonstrate here that not only is this
the case, but that accuracies are comparable to some of the older ab initio
calibration work in Table II.

Computed DFT electron affinities are compared with the best
nonrelativistic ab initio values in Table V, while basis set
convergence in the DFT results is depicted in Table VI for
two representative DFT functionals, one ``pure'', the other
hybrid.

As seen in Table VI, basis set convergence for the DFT results is
quite rapid. Convergence is essentially achieved from AVTZ basis
sets onwards, and extrapolations of any kind would add little to
the quality of the results. In the remainder of our discussion,
we will therefore employ the unextrapolated results with the
largest basis set, AV5Z.

The most striking feature about Table V is that performance with
many of the functionals is qualitatively different for first-row and
second-row atoms. 
As could be expected, the worst performance is put in by LDA, with a global mean
absolute deviation of 0.377 eV; however, the results for this functional are substantially better 
for second-row than for first-row atoms, 
the performance being almost as good as for the BP86 functional.  
Upon closer inspection (as exemplified
by comparison of the BPW91, B3PW91, and G96PW91 results), it seems
that the results for the first row (aside from hydrogen) are quite
sensitive to the nature of the exchange functional, while this is
much less the case for the second-row atoms, where the results
are rather dominated by the correlation functional.
The performance of many of the exchange-correlation functionals 
however for the simplest of
systems, i.e. the hydrogen atom, leaves a lot to be desired.
Considering first the ``pure DFT'' (nonhydrid) exchange-correlation
functionals as a group, it appears that the PW91 correlation
functional performs somewhat better than its LYP counterpart,
particularly for the second row. For exchange B88 works somewhat
better than PW91 for the first row, although there seems to be
little to choose between them for the second row. The 1996 Gill
exchange functional however appears to be markedly superior to both
of them, the differences again being most conspicuous for the
first row.  Compared to G96LYP, the different correlation functional
in G96PW91 cuts the error for the second row in half even as the
overall performance for the first row is comparable to that of
B3LYP. Overall G96PW91 emerges as the best ``pure DFT'' functional
for the criterion used here, with a mean absolute error of 0.11 eV
for atomic electron affinities (only 0.06 eV in the second row).
The contention that the PW91 correlation functional is best used in
conjunction with the PW91 exchange functional does not appear to
be borne out by the present results.

Turning now to the hybrid functionals, we note that the popular B3LYP
functional in fact performs slightly less well than G96PW91.
Performance for B3PW91 is in fact markedly better than that of B3LYP,
and the best of all the pre-1996 functionals considered. In line
with the general observation that the first-row EAs appear to
be much more sensitive to the exchange part of the functional than
their second-row counterparts, the admixture of Hartree-Fock exchange
also has the largest effect for the first row.

Interestingly, the 1-parameter B1LYP represents a dramatic
improvement over the 3-parameter B3LYP for first-row atoms. In
fact, its performance for the first-row electron affinities is not
dissimilar from some of the ab initio calibration studies in the
past. Performance for the second row is marred by a particularly
poor result for Si. LG1LYP yields marginally better results than
B1LYP for the second-row atoms, but slightly worse ones (on
average) for the first row. The mPW1LYP functional, on the
other hand, exhibits a slight performance improvement over
B1LYP for both first-and second-row atoms: residual errors for
the first row are down to +0.02 eV (H), +0.05 eV (B), -0.06 eV (C),
+0.03 eV (O), and -0.11 eV (F). Again the weakest performance
for the second row is put in for Si (-0.21 eV).

Interestingly enough, substitution of the PW91 correlation functional
leads to a serious deterioration of results for the first-row atoms:
this is perhaps to some extent related to the fact that the LYP
correlation functional was itself based on a fit\cite{Col75}
to estimated correlation
energies for the first-row atoms. The mPW1PW91 functional, on the
other hand, yields very good results for the second-row atoms,
with residual errors of -0.10 eV (Al), +0.04 eV (Si), +0.07 eV (P),
-0.01 eV (S), and -0.08 eV (Cl).

The fact that mPW1LYP seems to put in the best performance for the
first row and mPW1PW91 for the second row naturally leads to the
suggestion that perhaps a hybrid of the two correlation functionals
may lead to the best results overall. If we were to assume that the
Kohn-Sham orbitals do not differ greatly between the mPW1LYP and mPW1PW91
approaches, then the ``optimum hybrid'' could be determined by
minimizing the mean absolute error of a linear combination
$a EA_{\rm mPW1LYP}+ (1-a) EA_{\rm mPW1PW91}$ in terms of $a$.  
This procedure shows some similarity with the 
``empirical density
functionals'' recently proposed by Pople and coworkers\cite{Ada98}
As it happens, we find the `optimum' value of $a$ to be 0.669, or
almost exactly 2/3. This yields an overall mean absolute error of
0.07 eV, of which 0.05 eV for first-row atoms and 0.08 eV for
second-row atoms. Individual errors are:
H -0.02, B +0.10, C +0.01, O +0.00, F -0.13, Al
-0.04, Si -0.13, P +0.08, S -0.02, and Cl -0.12 eV --- in fact
the value for O accidentally agrees with the ab initio value to
four figures.
The present results can be considered as very promising for the
accurate calculation of electron affinities of large molecular systems.

\section{Conclusions}

We have carried out ab initio calibration calculations of the
electron affinities of the first-and second-row atoms. Our calculations
include extrapolations to the infinite-basis limit as well as corrections
for scalar relativistic and spin-orbit effects. Our best ab initio
values agree with the most recent experimental values to within better
than 0.001 eV on average. Neglect of correlation effects beyond CCSD(T)
causes an increase in the mean absolute error by an order of magnitude.
Inner-shell correlation is most important for the early second-row
elements, while scalar relativistic effects are quite important for the
later second-row elements. Neglect of electron correlation effects
on the scalar relativistic contributions leads to significant overestimates,
while inclusion of subvalence correlation is essential for accurate
spin-orbit splitting constants for the second-row elements.

The DFT results are essentially converged with respect to
extension of the basis set at the AVTZ level. The performance of DFT
methods for the 1st-row atoms is very strongly dependendent on the
quality of the exchange functional, while this is not
the case for second-row atoms where the correlation functional
appears to be rather more important. While the LYP correlation
functional works best for first-row atoms, its PW91 counterpart
appears to be preferable for second-row atoms. Among ``pure DFT'' (nonhybrid)
functionals, G96PW91 (Gill 1996 exchange combinated with Perdew-Wang
1991 correlation) puts in the best overall performance, actually slightly
better than the popular hybrid B3LYP functional. B3PW91 outperforms
B3LYP, while the recently proposed 1-parameter hybrid functionals
such as B1LYP
appear to be clearly superior to B3LYP and B3PW91 for first-row
atoms. mPW1LYP puts in the overall best performance for first-row
atoms, while mPW1PW91 yields the best results for second-row
atoms. The best overall performance appears to be afforded by
an empirical superposition of these functionals,
(2/3)mPW1LYP + (1/3)mPW1PW91.

\acknowledgments

JM is a Yigal Allon Fellow, the incumbent of the Helen and Milton
A. Kimmelman Career Development Chair (Weizmann Institute), and
an Honorary Research Associate (``Onderzoeksleider
in eremandaat'') of the
National Science Foundation of Belgium (NFWO/FNRS). GdO acknowledges
the Feinberg Graduate School (Weizmann Institute)
for a Postdoctoral Fellowship.
This research was supported by the Minerva Foundation, Munich,
Germany (JM).

\begin{table}
\caption{Best ab initio computed electron affinities (eV)}
\squeezetable
\begin{tabular}{lcccccccc}
 & SCF limit & CCSD(T) limit & core corr & spin-orbit & Darwin+MV & FCI corr. & Best calc. & Expt.[*]\\
 & $A+B.C^n$ & $A+B/n^3$ & CCSD(T)/ & CAS-CI(all)/ & ACPF(all)/ACVQZ & see & &\\
 &           &         &MTav5z  &MTav5z&MTav5z&text&\\
 \hline
H & -0.32877 & 1.08297 & 0.00000 & 0.00000 & -0.00004 & 0.00000 & 0.75416 & 0.754195\\
B & -0.26754 & 0.52465 & 0.00427 & -0.00060 & -0.00127 & 0.01907 & 0.27858 & 0.277(10),
0.279723(25)\cite{Sch98B}\\
C & 0.54826 & 0.70047 & 0.00720 & -0.00332 & -0.00283 & 0.01309 & 1.26288 & 1.2629(3)\\
N & -- & -- & -- & -- & -- & -- & -- & --\\
O & -0.53902 & 1.99391 & 0.00173 & -0.00222 & -0.00588 & 0.01223 & 1.46075 & 1.461122(3)\\
F & 1.30727 & 2.11864 & 0.00430 & -0.01652 & -0.00928 & 0.00056 & 3.40496,3.40285a & 3.401190(4)\\
\hline
Al & 0.04101 & 0.40219 & -0.01617 & -0.00385 & -0.00536 & 0.01497 & 0.43277 & 0.441(10),
0.43283(5)\cite{Sch98Al},\\
   &         &         &          &          &          &         &         & 0.44094(66)\cite{Cal96}\\
Si & 0.95579 & 0.46046 & -0.00965 & -0.01806 & -0.00787 & 0.00992 & 1.39060 & 1.385(5), 1.38946(6)\cite{Tho96}\\
P & -0.45796 & 1.19166 & -0.00521 & 0.01229 & -0.00937 & 0.01124 & 0.74264,0.74474b & 0.7465(3)\\
S & 0.90388 & 1.18400 & -0.00161 & -0.00410 & -0.01223 & 0.00441 & 2.07436,2.07544b & 2.077104(1)\\
Cl & 2.52999 & 1.13398 & 0.00085 & -0.03657 & -0.01509 & -0.00309 & 3.61008, 3.61113b & 3.61269(6), 3.612641(27)c\\
\end{tabular}

[*] Ref.\cite{CRC78} unless indicated otherwise

(a) including $A+B/l^3$ extrapolation of CCSDT--CCSD(T) difference
from AVTZ and AVQZ basis sets (see text)

(b) using dAVQZ and dAV5Z basis sets for valence correlation
extrapolation

(c) U. Berzinsh, M. Gustafsson, D. Hanstorp, A. Klinkm\"uller, U. Ljungblad, and A.-M. M{\aa}rtensson-Pendrill,  \PRA{51}{231}{1995}.

\end{table}

\begin{table}
\caption{Comparison of presently computed ab initio electron affinities (eV) with
earlier calculations}
\squeezetable
\begin{tabular}{lcllllcccccc}
 & Year & Source & Ref. & Level of theory & Basis set & H & B & C & O & F\\
 \hline
R & 1999 & This Work &  & Best calc. &  & 0.7542 & 0.2786 & 1.2629 & 1.4607 & 3.4029\\
R &  & \multicolumn{3}{l}{Most recent experimental values} &  & 0.7542(2) & 0.279723(25) & 1.2629(3) & 1.461122(3) & 3.401190(4)\\
NR & 1999 & Gdanitz & \cite{Gda99} & r12-MRACPF & (a) & 0.7542 & 0.2833 & 1.2655 & 1.454 & 3.398\\
R  & 1999 & Gdanitz & \cite{Gda99} & r12-MRACPF & (a,f) & 0.7538 & 0.2820 & 1.2623 & 1.445 & 3.385\\
NR & 1998 & Gutsev et al. & \cite{Gut98} & CCSDT & AV5Z & 0.747 & 0.241 & 1.259 &  1.432 & 3.395\\
SO  & 1998 & Curtiss et al. & \cite{g3} & G3 theory &  &  & 0.204 & 1.193 & 1.336 & 3.400\\
R & 1998 & Gou-xin et al. & \cite{Guo98} & LDA & Numer. & 0.637 & 0.282 & 1.220 & 1.292 & 2.180\\
R & 1997-8 & Wijesundera & \cite{Wij98,Wij97} & MC Dirac-Fock & Numer. & &0.260 & 1.210   &  & \\
R & 1997 & Eliav et al. & \cite{Eli97} & CCSD (+T) & (c) & &0.279 &   &  & \\
R & 1995 & Fischer et al. & \cite{Fis95} & MCHF+core+val. & Numer. & &0.2795 &   &  & \\
R & 1993 & Hughes \& Kaldor & \cite{Hug93} & Fock space CCSD & 13s9p6d4f2g &  &  &  &  & 3.421\\
NR & 1992-3 & Dunning et al. & \cite{Ken92} & FCIapprox & AVQZ & 0.740 & 0.263 & 1.246 & 1.401 & 3.364\\
NR & 1992 & Strout \& Scuseria & \cite{Str92} & CCSD(T) & 23s26p10d5f3g &  &  &  & 1.415 & \\
NR & 1992 & Moskowitz\&Schmidt & \cite{Mos92} & \multicolumn{2}{c}{Variational QMC} &    & 0.24(2) & 1.27(2) & 1.30(2) & 3.46(4) \\
NR & 1991 & Noro et al. & \cite{Nor91} & MRCI & 13s11p6d5f5g5h &  & 0.278 & 1.264 & 1.454 & 3.363\\
R & 1990 & Sundholm \& Olsen & \cite{Sun90} & (d) & Numer. &  & 0.2668 &  &  & \\
NR & 1989 & Novoa et al. & \cite{Nov91} & CIPSI-3 & 7s6p4d2f &  & 0.28 & 1.22 & 1.23 & 3.16\\
(e) & 1986 & Bauschlicher et al. & \cite{Bau86a,Bau86b} & FCI & (b) &  &  &  & 1.287 & 3.040\\
NR & 1985 & Feller \& Davidson & \cite{Fel85} & MR-CI+Q &  &  &  & 1.235 & 1.405 & \\
NR & 1985 & Raghavachari & \cite{Rag85} & CCD+ST(CCD) & 7s5p4d2f &  & 0.22 & 1.22 & 1.36 & 3.35\\
\hline
 &  &  &  &  &  & Al & Si & P & S & Cl & \\
\hline
R &  & This Work &  & Best calc. &  & 0.4328 & 1.3906 & 0.7467 & 2.0768 & 3.6111 & \\
R &  &\multicolumn{3}{l}{Most recent experimental values}&  & 0.43283(5) & 1.38946(6) & 0.7465(3) & 2.077104(1) & 3.61264(3) & \\
NR & 1998 & Gutsev et al. & \cite{Gut98} & CCSDT & AV5Z & 0.433 & 1.405 & 0.714 & 2.059 & 3.623 & \\
R  & 1998 & Curtiss et al. & \cite{g3} & G3 theory &  & 0.390 & 1.379 & 0.711 & 2.064 & 3.608 & \\
R & 1998 & Gou-xin et al. & \cite{Guo98} & LDA & Numer. & 0.450 & 1.372 & 0.748 & 1.996 & 3.332 & \\
R & 1997-8 & Wijesundera & \cite{Wij98,Wij97} & MC Dirac-Fock & Numer. & 0.433 &  & 0.702 &  &  & \\
R & 1997 & Eliav et al. & \cite{Eli97} & CCSD (+T) & (c) & 0.427 &  &  &  &  & \\
NR & 1996 & Greeff et al. & \cite{Gre96} & \multicolumn{2}{c}{Diffusion QMC} & 0.432(21) &\\
R & 1995 & Heinemann et al. & \cite{Hei95} & CCSD(T)$^g$ & 6s5p4d3f3g2h1i &   &  &  &
2.064(5) &\\
R & 1993 & Hughes \& Kaldor & \cite{Hug93} & FS:CCSD & 13s9p6d4f2g &  &  &  &  & 3.608 \\
SO & 1992-3 & Dunning et al. & \cite{Woo93} & FCIapprox & AVQZ & 0.441 & 1.413 & 0.702 & 2.051 & 3.632 \\
NR & 1988 & Yoshida et al. & \cite{Yos88} & \multicolumn{2}{c}{Diffusion QMC+ECP} &    &   &   &   & 3.617(198) \\
\end{tabular}

The prefixes NR and R indicate nonrelativistic and relativistic values/results, respectively; SO indicates values with only a correction for spin-orbit
splitting applied (not for scalar relativistic effects)

(a) 19s14p7d5f3g2h(B,C)  19s14p8d6f4g3h1i(O,F)  11s5p4d3f2g(H)

(b) O: FCI(2p only)/[6s5p3d2f]; F: FCI(full valence)/[5s4p2d]

(c) 35s26p20d14f9g6h4i

(d) MCHF + core-valence + relativistic corrections

(e) R for oxygen, NR for fluorine

(f) Relativistic corrections taken from numerical HF calculations,
Ref.\cite{Gar95}

(g) Plus $n$-particle correction from valence FCI in $[5s4p2d1f]$ basis
set (+0.005 eV); scalar relativistic (DMV) contribution from MRCI+Q/6s5p4d3f2g calculation (-0.008 eV);
spin-orbit from experiment (-0.004 eV).

\end{table}

\begin{table}
\caption{Effect of electron correlation on the computed scalar
relativistic corrections (eV)}
\squeezetable
\begin{tabular}{lcccccc}
 & SCF & ACPF/ & SCF & ACPF/ & Num. HF & Num. HF\\
 & MTavqz & MTavqz & MTav5z & MTav5z & \cite{Tat97} & \cite{Gar95}\\
\hline
H & -0.00010 & -0.00004 & -0.00010 & -0.00004 & -0.00016 & +0.0000\\
B & -0.00144 & -0.00128 & -0.00143 & -0.00127 & -0.00148 & -0.0013\\
C & -0.00323 & -0.00283 & -0.00323 & -0.00283 & -0.00345 & -0.0032\\
O & -0.00796 & -0.00592 & -0.00795 & -0.00588 & -0.00819 & -0.0080\\
F & -0.01236 & -0.00930 & -0.01236 & -0.00928 & -0.01319 & -0.0129\\
\hline
Al & -0.00528 & -0.00536 & -0.00529 & -0.00536 & -0.00538 & -0.0054\\
Si & -0.00892 & -0.00786 & -0.00892 & -0.00787 & -0.00922 & -0.0092\\
P & -0.01056 & -0.00935 & -0.01056 & -0.00937 & -0.01087 & -0.0109\\
S & -0.01426 & -0.01219 & -0.01427 & -0.01223 & -0.01473 & -0.0147\\
Cl & -0.01830 & -0.01504 & -0.01831 & -0.01509 & -0.01917 & -0.0192\\
\end{tabular}

\end{table}

\begin{table}
\caption{Effect of dynamical correlation on the computed
atomic spin-orbit fine structures (cm$^{-1}$). Degeneracies are
given in parentheses with the experimental values}
\squeezetable
\begin{tabular}{llccccc}
&  & MTavtz & MTav5z & MTav5z & MTav5z & MTav5z\\
&Expt.[*] & +CI(all)$^c$ & CASSCF & +CI(val)$^a$ & +CI(subval)$^b$ & +CI(all)$^c$\\
\hline
B & 0(2), 15.254(4) & 14.4 & 14.9 & 14.2 &  & 14.7\\
B$^-$ & 0(1), 4(3), 9(5); 0(1), 3.23(3), 8.41(5)\cite{Sch98B} & 2.45,7.34 & 2.44,7.31 & 2.4,7.2 &  & 2.49,7.50\\
C & 0(1), 16.40(3), 43.40(5) & 13.2,39.5 & 13.1,39.2 & 13.2,39.6 &  & 13.4,40.2\\
O & 0(5), 158.265(3), 226.977(1) & 153.2,229.8 & 161.0,241.6 & 153.6,230.4 &  & 155.1,232.6\\
O$^-$ & 0(4), 177.08(2) & 176.8 & 180.6 & 177.2 &  & 179.0\\
F & 0(4), 404.1(2)  & 394.9 & 401.9 & 397.7 &  & 399.8\\
Al & 0(2), 112.061(4)  & 114.6 & 103.2 & 90.0 & 115.0 & 115.4\\
Al$^-$ & 0(1), 26.0(3), 76.0(5); 0(1), 22.7$\pm$0.3(3), 68.4$\pm$0.4(5)\cite{Sch98Al} & 22.8,68.3 & 19.6,58.9 & 18.6,55.7 & 22.9,68.8 & 22.9,68.8\\
Si & 0(5), 77.113(3), 223.157(1) & 72.7,218.0 & 63.8,191.3 & 61.7,185.1 & 72.8,218.5 & 72.8,218.5\\
P$^-$ & 0(5), 181(3), 263(1) & 197.7,296.5 & 180.51,270.76 & 165.9,248.7 & 197.3,296.0 & 198.2,297.3\\
S& 0(5), 396.09(3), 573.65(1) & 394.4,592.6 & 366.3,549.5 & 347.7,521.5 & 394.3,591.4 & 395.3,593.0\\
S$^-$ & 0(4), 483.54(2) & 492.8 & 452.9 & 436.3 & 492.7 & 493.8\\
Cl & 0(4), 882.36(2) & 883.1 & 823.2 & 799.3 & 883.5 & 884.8\\
\end{tabular}

[*] Taken from Ref.\cite{Jan85} unless indicated otherwise. Note that the
degeneracies for Si($^3P_n$) ($n$=0,1,2) in Ref.\cite{Jan85} are
misprinted.

(a) CASSCF-CI, only excitations from valence orbitals considered

(b) CASSCF-CI, excitations from valence orbitals as well as (2s2p)
sub-valence orbitals considered

(c) CASSCF-CI, also including excitations from (1s) orbitals

\end{table}

\begin{table}
    \caption{Performance of different exchange-correlation functionals
    for atomic electron affinities (eV). The AV5Z basis set was used
    throughout}
\squeezetable
    \begin{tabular}{lcccccccc}
&\multicolumn{8}{c}{Nonhybrid functionals}\\
  & Best   & LDA & B-    & B-    & B-    & PW91- & G96-  & G96-\\
  & nonrel.$^{a}$&     & LYP   & P86   & PW91  & PW91  & LYP   & PW91\\
    \hline
H  & 0.75420& 0.952 & 0.881 & 1.037 & 0.760 & 0.767 & 0.841 & 0.721\\
B  & 0.28045& 0.756 & 0.468 & 0.701 & 0.605 & 0.659 & 0.422 & 0.557\\
C  & 1.26903& 1.814 & 1.367 & 1.646 & 1.562 & 1.630 & 1.325 & 1.519\\
O  & 1.46885& 2.071 & 1.839 & 1.918 & 1.726 & 1.841 & 1.719 & 1.604\\
F  & 3.43077& 4.128 & 3.681 & 3.759 & 3.601 & 3.724 & 3.592 & 3.511\\
Al & 0.44199& 0.646 & 0.390 & 0.657 & 0.569 & 0.610 & 0.356 & 0.535\\
Si & 1.41653& 1.593 & 1.231 & 1.552 & 1.473 & 1.521 & 1.202 & 1.444\\
P  & 0.73973& 1.039 & 0.911 & 1.076 & 0.870 & 0.931 & 0.842 & 0.802\\
S  & 2.09069& 2.393 & 2.129 & 2.310 & 2.136 & 2.216 & 2.060 & 2.068\\
Cl & 3.66173& 3.929 & 3.571 & 3.765 & 3.617 & 3.702 & 3.515 & 3.562\\
\hline
\multicolumn{2}{l}{Mean abs. error} 
             &  0.377 &     0.157 &  0.287&  0.145&  0.205&  0.128&  0.108\\
\multicolumn{2}{r}{First row} 
             & 0.504 &     0.206 & 0.372 & 0.210 & 0.283 & 0.139 & 0.155\\
\multicolumn{2}{r}{Second row} 
             & 0.250 &     0.108 & 0.202 & 0.081 & 0.126 & 0.116 & 0.061\\
\hline
&\multicolumn{8}{c}{Hybrid functionals}\\
\hline
  &  Best  &  B3-  & B3-  &  B1-  &  LG1- &  mPW1-&  mPW1-&  (b)\\
  & nonrel.$^{a}$&  LYP  & PW91 &  LYP  &  LYP  &  PW91 &  LYP  &  \\
     \hline
H  & 0.75420& 0.926 & 0.761 & 0.765 & 0.827 & 0.659 & 0.770 & 0.733 \\
B  & 0.28045& 0.476 & 0.522 & 0.308 & 0.352 & 0.487 & 0.332 & 0.383 \\
C  & 1.26903& 1.380 & 1.470 & 1.183 & 1.213 & 1.425 & 1.211 & 1.282 \\
O  & 1.46885& 1.688 & 1.509 & 1.454 & 1.583 & 1.403 & 1.502 & 1.469 \\
F  & 3.43077& 3.527 & 3.376 & 3.270 & 3.347 & 3.253 & 3.318 & 3.297 \\
Al & 0.44199& 0.466 & 0.551 & 0.324 & 0.342 & 0.539 & 0.341 & 0.407 \\
Si & 1.41653& 1.345 & 1.473 & 1.184 & 1.192 & 1.463 & 1.204 & 1.290 \\
P  & 0.73973& 0.964 & 0.854 & 0.804 & 0.859 & 0.806 & 0.831 & 0.823 \\
S  & 2.09069& 2.203 & 2.128 & 2.029 & 2.070 & 2.083 & 2.063 & 2.069 \\
Cl & 3.66173& 3.672 & 3.626 & 3.490 & 3.517 & 3.583 & 3.524 & 3.544 \\
\hline
\multicolumn{2}{l}{Mean abs. error} &
              0.124& 0.090 & 0.095 & 0.101 & 0.100 & 0.084 & 0.065\\
\multicolumn{2}{r}{First row} &
              0.159& 0.109 & 0.060 & 0.080 & 0.140 & 0.054 & 0.054\\
\multicolumn{2}{r}{Second row} &
              0.088& 0.071 & 0.130 & 0.122 & 0.059 & 0.114 & 0.077\\
\end{tabular}

(a) this work: best ab initio minus spin-orbit and scalar relativistic
contributions

(b) (2/3) mPW1LYP + (1/3) mPW1PW91

\end{table}

\begin{table}
    \caption{Basis set convergence of computed electron affinities
    (eV) for selected exchange-correlation functionals}
    \squeezetable
\begin{tabular}{lcccccccc}
   &\multicolumn{4}{c}{G96PW91/}&\multicolumn{4}{c}{mPW1LYP/}\\
&    AVDZ&   AVTZ &  AVQZ  & AV5Z   &AVDZ    &AVTZ    &AVQZ    &AV5Z\\
\hline
H  & 0.6825 & 0.7019 & 0.7054 & 0.7210 & 0.7270 & 0.7504 & 0.7550 & 0.7703\\
B  & 0.5289 & 0.5376 & 0.5422 & 0.5565 & 0.3064 & 0.3123 & 0.3205 & 0.3317\\
C  & 1.5097 & 1.5141 & 1.5132 & 1.5194 & 1.1998 & 1.2004 & 1.2050 & 1.2113\\
O  & 1.5914 & 1.5929 & 1.5950 & 1.6041 & 1.4921 & 1.4910 & 1.4950 & 1.5015\\
F  & 3.5496 & 3.5189 & 3.5096 & 3.5106 & 3.3590 & 3.3216 & 3.3170 & 3.3184\\
Al & 0.5372 & 0.5374 & 0.5315 & 0.5346 & 0.3342 & 0.3341 & 0.3381 & 0.3412\\
Si & 1.4580 & 1.4555 & 1.4434 & 1.4439 & 1.2094 & 1.2060 & 1.2035 & 1.2042\\
P  & 0.7601 & 0.8025 & 0.7961 & 0.8021 & 0.7932 & 0.8285 & 0.8268 & 0.8312\\
S  & 2.0850 & 2.0751 & 2.0668 & 2.0676 & 2.0747 & 2.0642 & 2.0623 & 2.0626\\
Cl & 3.6204 & 3.5749 & 3.5637 & 3.5621 & 3.5769 & 3.5301 & 3.5259 & 3.5244\\
\end{tabular}
\end{table}

\end{document}